\documentclass[a4paper, 10 pt, conference]{ieeeconf}

\usepackage{calrsfs,times}
\usepackage{graphicx,psfrag,enumerate}
\usepackage{amssymb}
\usepackage{amsmath}
\usepackage{calrsfs}
\usepackage{color, epsfig, epstopdf, wrapfig}
\usepackage{paralist}
\usepackage{float,cuted}

\newtheorem{theorem}{Theorem}
\newtheorem{remark}{Remark}

\usepackage{amsmath}
\usepackage{accents}
\newlength{\dhatheight}

\usepackage{color}
\usepackage[normalem]{ulem}
\usepackage{soul} 
\soulregister\cite7
\soulregister\ref7
\soulregister\pageref7

\IEEEoverridecommandlockouts

\begin{document}

\title{\LARGE \bf 
  Robust 
  Detection of Biasing
  Attacks on Misappropriated Distributed Observers via Decentralized
  $H_\infty$ synthesis    
  \thanks{This work was supported by the Australian
  Research Council and the University of New South Wales.}}

\author{V.~Ugrinovskii\thanks{V. Ugrinovskii is with the School of Engineering and Information Technology, University of New South Wales at the Australian Defence Force Academy, Canberra, ACT 2600, Australia. {\tt\small v.ougrinovski@adfa.edu.au}}}

\maketitle
         
\begin{abstract}
We develop 
a decentralized $H_\infty$ synthesis 
approach to detection of 
biasing misappropriation attacks on distributed observers. Its starting
point is to equip the observer with an attack model 
which is then used in the design of attack 
detectors. A two-step design procedure  is
proposed. First, an initial centralized setup is carried out which 
enables each node to compute the parameters of its attack detector
online in a decentralized manner, without interacting with other
nodes. Each such detector is designed using the 
$H_\infty$ approach. 
Next, the attack detectors are embedded into the network, which allows them
to detect misappropriated nodes from 
innovation in the network interconnections. 
\end{abstract}

\section{Introduction}

The need to protect control systems from 
malicious 
attacks has led to an emergence of methodologies for resilient
control. Resilient control schemes intend to enhance tolerance of control
systems to attacks. 
The interest in the
resilience problem has increased substantially after situations were
discovered where an adversary was able to compromise integrity of a
control system by injecting a malicious input into the measurements, which
was not detected by bad data detection units~\cite{DS-2010a}. 

Cooperative networked control systems are particularly vulnerable to such
attacks
. For instance, in networks of observers~\cite{GS-2011}
fidelity of information is crucial, it makes possible 
estimation of the plant even when it cannot be reliably estimated from local
measurements. A misappropriated observer node
can provide a false
 information to its neighbours, and use them to bias the entire
network. On the other hand, as this paper
shows, information sharing between network nodes provides opportunities for
monitoring integrity of distributed filter networks. 

This paper considers the problem of detection of malicious attacks,
known as biasing attacks~\cite{TSSJ-2015}, on distributed state
observers. The problem was posed originally in~\cite{DUSL1a}, it is motivated by
the necessity to enhance resilience properties of a general class of distributed
state estimation networks such as those introduced
in~\cite{Olfati-Saber-2007,SWH-2010,U6,LaU1}. It is concerned with a
situation where one of the observers in the network is misappropriated and
is used to supply a biased information to its neighbours. 

While we adopt the same bias injection attack model as in~\cite{DUSL1a} and 
are concerned with detecting the same biasing behaviour of misappropriated
nodes (also cf.~\cite{TSSJ-2014}),  our approach to the synthesis of
attack detectors is different from~\cite{DUSL1a}. It is based on the
decoupling technique developed in~\cite{ZU1a}. This allows us to dispense
with several difficulties 
of the vector dissipativity approach adopted
in~\cite{DUSL1a}. First and foremost, our approach provides the observer
nodes with a better computational autonomy. In this paper, the process of
computing the characteristics of an attack
detector at each node is independent of other nodes
. Although an initial centralized
setup is required for this, it involves only characteristics of the
communication network, and can be completed without knowledge of the system
and the filter characteristics. This contrasts our results with the technique
in~\cite{DUSL1a} where the design conditions are 
coupled. 
As a by-product of decentralization, our technique  
applies to more general time-varying distributed
filters. It also brings several other
improvements, which simplify tuning the
detector. At the same time, the proposed scheme retains advantages of
the attack detection scheme developed in~\cite{DUSL1a}. It is 
robust against uncertainties in the sensors and the plant model and relies
on the same sensory and interconnection data as the networked observer it
seeks to protect.  

The paper begins with presenting a biasing attack
model from~\cite{DUSL1a} and also gives a background on the distributed
filtering in Section~\ref{sec:distributed_estimation},
respectively. Section~\ref{Problem.formulation} presents the problem formulation
and our main results, which allow one to construct a collaborative attack
detector. The design technique is described in Section~\ref{design.sec}
. The conclusions are given in Section~\ref{sec:conclusion}.

\emph{Notation}: $\mathbf{R}^n$ denotes the real Euclidean $n$-dimensional vector space, with the norm  $\|x\|=(x'x)^{1/2}$; here the symbol $'$ denotes the transpose of a matrix or a vector.
The symbol $I$ denotes the identity matrix. For real symmetric
$n\times n$ matrices $X$ and $Y$, $Y>X$ (respectively, $Y\geq X$) means the
matrix $Y-X$ is positive definite (respectively, positive semidefinite). 
The notation $L_2[0, \infty)$ refers to the Lebesgue space of
$\mathbf{R}^n$-valued vector-functions $z(.)$, defined on the time interval
$[0, \infty)$, with the norm $\|z\|_2\triangleq\left(\int_0^\infty
  \|z(t)\|^2 dt \right)^{1/2}$ and the inner product $\int_0^\infty z_1'(t)
z_2(t) dt$. 
   
\section{Biasing misappropriation attacks on distributed observers}
\label{sec:distributed_estimation}

A distributed observer problem under consideration involves estimation of the state of a
time varying plant 
\begin{eqnarray}
\label{state}\label{eq:plant}
 \dot{x} = A(t)x +B(t)w, \quad x(0)=x_0,
\end{eqnarray}
subject to an unknown modeling disturbance $w$, 
from a collection of measurements 
\begin{equation}\label{measurement}
y_i = C_i(t)x + D_i(t)v_i, \quad i=1,2,\ldots,N,
\end{equation}
taken at $N$ nodes of a sensor network, and also affected by measurement
disturbances $v_i$. The distributed estimation problem requires that the
measurements are to be processed at the sensor nodes, rather than
centrally. For this, sensors-observers are interconnected into a network,
so that the constituent nodes can share their pre-processed data and can
enhance their local estimates of the plant state. 

To describe the problem mathematically, let us assume that
the state $x$ and the disturbance $w$ are respectively in $\mathbb{R}^{n}$,
$\mathbb{R}^{m}$, and each measurement $y_i$ is in $\mathbb{R}^{p_i}$. The
disturbances $w$ and $v_i\in\mathbb{R}^{m_i}$ will be assumed 
$\mathcal{L}_2$ integrable on $[0,\infty)$. The initial state $x_0$ is also
unknown. A typical distributed estimation problem involves constructing a
network of filters of the form
\begin{eqnarray}
\dot{\hat{x}}_i &=& A(t)\hat x_i + L_i(t)(y_i-C_i(t)\hat x_i) \nonumber \\
&&+\sum_{j\in\mathbf{N}_i}K_{ij}(t)(c_{ij}-W_{ij}\hat x_i), \quad
\hat{x}_i(0)=\xi_i, 
\label{filter_i}\label{UP7.C.d.unbiased}
  \end{eqnarray}
each generating an estimate $\hat x_i(t)$ of the plant state $x(t)$ using 
its measurement $y_i$ and the information received
from the neighbours, in the form of a $p_{ij}$-dimensional signal 
\begin{equation}\label{communication}
c_{ij} = W_{ij}\hat{x}_j + H_{ij}v_{ij},\quad j\in\mathbf{N}_i;
\end{equation}
here $\mathbf{N}_i$ denotes the neighbourhood of agent $i$, i.e., the set
of network nodes that communicate their information to $i$. Since the plant
is time-varying, the filter coefficients $L_i$, $K_{ij}$ are allowed to be
time-varying as well. 

The matrices $W_{ij}$ determine which part of the vector $\hat x_j$ is made
available to node $i$ by node $j$. Typically, distributed observers are required
where the plant may not be detectable from local measurements at some of
the nodes, 
and the signals $c_{ij}$ serve to complement the local 
measurements with an additional information. Mathematically, to obtain an
unbiased estimate of $x$, node $i$ may require a portion of $\hat x_j$ which
lies in the subspace of states undetectable from $y_i$. The
matrix $W_{ij}$ can be thought of as `projecting' $\hat x_j$ onto that
subspace. However, the communication channels which deliver this information are
typically subject to disturbances; this is reflected in the term
$H_{ij}v_{ij}$ in (\ref{communication}). 

The problem is to determine
estimator gains $L_i$ and $K_{ij}$ in \eqref{UP7.C.d.unbiased} to ensure
that each estimate $\hat x_i$ (or a part of it) converges to $x(t)$ in some
sense, and that some filtering performance against
disturbances is guaranteed. On the other hand, an adversary may seek
to prevent this from happening. A common means for interfering with
a normal operation of the system involve injecting false data into
the measurements or communications~\cite{PDB-2013}, however consideration is
also given to misappropriation attacks where adversary gains control over the
control algorithm and modifies it according to its strategic
goals~\cite{Smith-2015}. In line with this idea, in~\cite{DUSL1a} we considered
a situation where the attacker interferes with dynamics of the hijacked
node by injecting a biasing input into the filter. In this
paper we extend this model to consider a similar type of attack on the
network of time-varying observers (\ref{UP7.C.d.unbiased}). Specifically,
we consider the situation where the
adversary substitutes one or several observers (\ref{UP7.C.d.unbiased})
with  
\begin{eqnarray}  
    \dot{\hat x}_i&=&A(t)\hat x_i + L_i(t)(y_i(t)-C_i(t)\hat x_i) \nonumber \\
&& +\sum_{j\in
      \mathbf{N}_i}K_{ij}(t)(c_{ij}-W_{ij}\hat x_i)+F_if_i, \quad \hat
    x_i(0)=\xi_i,\quad 
  \label{UP7.C.d}
\end{eqnarray}
Here $F_i\in\mathbf{R}^{n\times n_{f_i}}$ is a constant matrix and $f_i\in \mathbf{R}^{n_{f_i}}$ is the unknown signal representing an attack
input. In the following, we will present an algorithm for detecting and
tracking these unknown inputs. 

\begin{figure}[t]
\psfrag{R}{$f_i$}
\psfrag{nu}{$-\nu_i$}
\psfrag{Y}{$\hat f_i$}
\psfrag{H}{$G_i(s)$}
\psfrag{G}{$\frac{1}{s}$}
\psfrag{+}{$+$}
\psfrag{-}{$-$}
  \centering
  \includegraphics[width=0.45\textwidth]{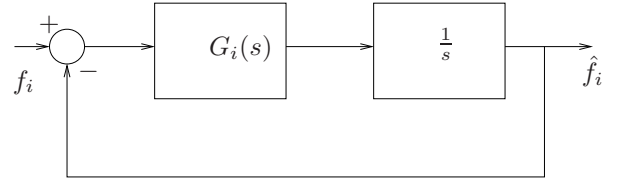}
  \caption{An auxiliary `input tracking' model.}
  \label{tracker}
\end{figure}

Following~\cite{DUSL1a}, we will consider a class of
attacks on the filter (\ref{UP7.C.d}) consisting of biasing
inputs $f_i(t)$ of the form
\begin{equation}
  \label{decomp}
f_i(t)=f_{i1}(t)+f_{i2}(t),  
\end{equation}
where the Laplace transform of $f_{i1}(t)$, $f_{i1}(s)$, is such that
$\sup_{\omega}|\omega f_{i1}(j\omega)|^2<\infty$ and
$f_{i2}\in L_2[0,\infty)$. It can be shown that one can select a proper
$n_{f_i}\times n_{f_i}$ transfer function $G_i(s)$ for which the system in 
Fig.~\ref{tracker} is stable and has the property 
\begin{equation}\label{f-eta} 
 \int_0^\infty\|f_i-\hat f_i\|^2dt<\infty, 
\end{equation}
which holds for all inputs that admit decomposition~(\ref{decomp}). 
In particular, bias injection attack inputs of the form
`constant + an exponentially decaying transient' generated by a low pass
filter introduced in~\cite{TSSJ-2015} have the form (\ref{decomp}), and
therefore can be `tracked' using a system shown in Fig.~\ref{tracker}.  
Of course, in reality it is not possible to track covert attack
inputs. However, the model in Fig.~\ref{tracker} is useful in that it
allows us to associate with the class of biasing attack inputs a system
\begin{eqnarray}
&&\dot\epsilon_i = \Omega_i\epsilon_i+\Gamma_i \nu_i, \qquad
\epsilon_i(0)=0, \label{Om.sys.general} \\ 
&&\hat f_i= \Upsilon_i\epsilon_i, \nonumber
\end{eqnarray}
where $\nu_i=\hat f_i-f_i$ is an $L_2$-integrable input, according to
(\ref{f-eta}). 

In practice, the transfer function $G_i(s)$ must be selected by the
designer according to an anticipated behaviour of the attack inputs
$f_i(t)$. For example, for the mentioned class of bias injection attacks
of the form `constant + an exponentially decaying transient', one
can choose $G_i(s)$ to be a first order low pass filter,
$G_i(s)=\frac{g_i}{s+2\beta_i}I$. The corresponding coefficients of the system
(\ref{Om.sys.general}) then take the form  
\begin{equation}\label{omega}
\Omega_i=\left[\begin{array}{cc} 0 & I \\ 0 & -2\beta_i
    I\end{array}\right], \quad \Gamma_i=\left[\begin{array}{c} 0  \\
          -g_i\end{array}\right], \quad \Upsilon_i=[I~0].
\end{equation}

The choice of the transfer functions $G_i(s)$ and the corresponding
system (\ref{Om.sys.general}) will affect performance of the attack detectors to
be introduced later in the paper. For instance, in the above example,
selecting $\beta_i$, $g_i$ enables tuning the performance of the
attack detectors.

\section{Problem Formulation}\label{Problem.formulation}

First, let us summarize the information which the observer
nodes can utilize to detect an attack. Each node has two innovation
signals available,   
\begin{eqnarray}
  \zeta_i&=&y_i-C_i(t)\hat x_i \nonumber \\
         &=&C_i(t)(x-\hat x_i) + D_i(t)v_i, 
                                              \label{out.y} \\
  \zeta_{ij}&=&c_{ij}-W_{ij}\hat x_i.
                                              \label{out.c}
\end{eqnarray}
The innovation $\zeta_i$ symbolizes the new information contained in the
measurement acquired by node $i$, compared with its own prediction of
that information. Likewise, the innovation $\zeta_{ij}$ symbolizes the
new information contained in the message that node $i$ receives from its
neighbour $j\in\mathbf{N}_i$. Similar to~\cite{DUSL1a}, these innovations
will play an 
instrumental role in detecting a rogue behaviour of misappropriated
nodes. Both signals are readily available at node $i$; computing them only
requires the local measurement $y_i$ and the neighbour messages $c_{ij}$.
$j\in\mathbf{N}_i$, available at node $i$, along with $\hat x_i$; see (\ref{UP7.C.d}).

The problem of detecting a biasing attack of the form (\ref{decomp}) on the
observer network comprised of observers (\ref{UP7.C.d}) consists in designing a
network of attack detectors
\begin{eqnarray}
  \label{detector.general}
  \dot\mu_i&=&\mathcal{A}_d(t)\mu_i + L_{d,i}(t)(\zeta_i-W_{d,i}\mu_i)
  \nonumber \\
  && +\sum_{j\in 
    \mathbf{N}_i}K_{d,ij}(t)(\zeta_{ij}-W_{d,ij}(\mu_j-\mu_i)), \\
  \varphi_i&=&C_{d,i}\mu_i, \qquad
  \mu_i(0)=\mu_{i,0}, \nonumber 
\end{eqnarray}
which guarantee that 
\begin{eqnarray}
\int_0^{+\infty}\|\varphi_i-f_i\|^2dt <+\infty.
\label{convergence}
\end{eqnarray}
Here, $\mathcal{A}_d(t)$, $L_{d,i}(t)$,  $K_{d,ij}(t)$, $W_{d,i}$, 
$W_{d,ij}$, $C_{d,i}$ are matrix valued coefficients to be found.

According to (\ref{convergence}), the output of each detector subsystem
(\ref{detector.general}) is required to track $f_i$ in the
$\mathcal{L}_2$ sense. This makes these outputs suitable as indicators of
the biasing attack. Note that 
each detector relies on the
information received from its neighbouring nodes contained in the 
innovation signals $\zeta_{ij}$. 

\section{Design of attack 
  detectors}\label{design.sec}

Define the estimation error at node $i$, $e_i=x-\hat x_i$. It follows from
(\ref{eq:plant}), (\ref{UP7.C.d}) that the dynamics of this error are
described by
 \begin{eqnarray}
    \dot{e}_i&=&(A(t) - L_i(t)C_i(t)-\sum_{j\in
      \mathbf{N}_i}K_{ij}(t)W_{ij})e_i \nonumber \\ &+&\sum_{j\in
      \mathbf{N}_i}K_{ij}(t)W_{ij}e_j +B(t)w-L_i(t)D_i(t)v_i  \nonumber \\ & 
-& \sum_{j\in
      \mathbf{N}_i}K_{ij}(t)H_{ij}v_{ij} 
      -F_if_i, \qquad  e_i(0)=x_0-\xi_i. \quad \label{e} 
\end{eqnarray}
Using the representation of the attack input
$f_i=\Upsilon_i\epsilon_i-\nu_i$ and 
(\ref{Om.sys.general}), the input $f_i$ can be eliminated from
(\ref{e}). This will result in the following extended system:
\begin{eqnarray}
    \dot{e}_i&=&(A(t) - L_i(t)C_i(t)-\sum_{j\in
      \mathbf{N}_i}K_{ij}(t)W_{ij})e_i \nonumber \\ && +\sum_{j\in
      \mathbf{N}_i}K_{ij}(t)W_{ij}e_j -F_i\Upsilon_i\epsilon_i+B(t)w-L_i(t)D_i(t)v_i
    \nonumber \\ &&  
-\sum_{j\in
      \mathbf{N}_i}K_{ij}(t)H_{ij}v_{ij} 
      +F_i\nu_i, \qquad  e_i(0)=x_0-\xi_i,  \nonumber \\
      \dot \epsilon_i&=&\Omega_i \epsilon_i+\Gamma_i \nu_i \qquad
      \epsilon_i(0)=0. \label{e.ext.nu.om} 
\end{eqnarray}
Also, using (\ref{out.y}), (\ref{out.c}), define outputs of the
extended system (\ref{e.ext.nu.om}) as
\begin{eqnarray}
  \zeta_i&=&C_i(t)e_i + D_iv_i, 
                                              \label{out.y.1} \\
  \zeta_{ij}&=&-W_{ij}(e_j-e_i)+H_{ij}v_{ij}, \quad j\in \mathbf{N}_i. 
                                              \label{out.c.1}
\end{eqnarray}

These outputs are available at each node of
the network and will be used as inputs to an 
attack detector, whose
function is to estimate the state of the system (\ref{e.ext.nu.om}) and
generate an output $\varphi_i$ that converges to $\hat f_i$ and $f_i$
while attenuating the disturbances.

The observer which we propose below for estimating the combined state
$(e_i,\epsilon_i)$ of this system is a time-varying version of the
observer introduced in~\cite{DUSL1a}. This observer is as follows:
\begin{eqnarray}
    \dot{\hat{e}}_i&=&(A(t) - L_i(t)C_i(t)-\sum_{j\in
      \mathbf{N}_i}K_{ij}(t)W_{ij})\hat{e}_i \nonumber \\ 
& +& \sum_{j\in
      \mathbf{N}_i}K_{ij}(t)W_{ij}\hat e_j-F_i\Upsilon_i\hat\epsilon_i
    \nonumber \\ 
    &+&\bar L_i(\zeta_i-C_i(t)\hat{e}_i)+\sum_{j\in
      \mathbf{N}_i}\bar K_{ij}(t)(\zeta_{ij}-W_{ij}(\hat{e}_i-\hat{e}_j)), \nonumber \\
  \dot{\hat\epsilon}_i &=& 
\Omega_i \hat\epsilon_i +
    \check L_i(\zeta_i-C_i\hat{e}_i) \nonumber \\
&+&\sum_{j\in
      \mathbf{N}_i}\check K_{ij}(t)(\zeta_{ij}-W_{ij}(\hat{e}_i-\hat e_j)), \nonumber \\
\varphi_i&=& \Upsilon_i\hat{\epsilon}_i, 
\qquad\qquad \hat{e}_i(0)=0, \quad \hat\epsilon_i(0)=0.\label{ext.obs.nu.1.om} 
\end{eqnarray}
From now, our effort is directed towards finding a constructive method
for computing the coefficients of this observer. We seek to obtain the
gains $\bar L_i(t)$, $\bar K_{ij}(t)$, $\check L_i(t)$, $\check K_{ij}(t)$   
such that the output $\varphi_i$ of each filter 
(\ref{ext.obs.nu.1.om}) converges to $f_i$. This will establish the filters
(\ref{ext.obs.nu.1.om}) as biasing attack detectors. The inputs to the observer (\ref{ext.obs.nu.1.om}) are innovation signals
(\ref{out.y}), 
(\ref{out.c}). 

Despite the similarity between the filter (\ref{ext.obs.nu.1.om}) and 
the filter introduced in~\cite{DUSL1a} for the similar task,  
the time-varying nature of the problem under consideration requires us to
revisit the design method obtained in~\cite{DUSL1a}. The design conditions
proposed in~\cite{DUSL1a} involve solving certain coupled linear matrix
inequalities. Although similar inequalities can be derived for the
problem considered here, they will involve derivatives of the vector storage
functions and will therefore be \emph{differential} matrix inequalities,
with time-varying coefficients. Such inequalities are very difficult to
solve, even using centralized solver facilities. For this
reason, here we develop an alternative technique for computing the
coefficients of the observer (\ref{ext.obs.nu.1.om}). Although our task of 
selecting the gains $\bar L_i$, $\bar K_{ij}$, $\check L_i$, $\check
K_{ij}$ so that the observer (\ref{ext.obs.nu.1.om}) attenuates the
$\mathcal{L}_2$-integrable  disturbances $w$, $v_i$, $v_{ij}$, $\nu_i$,
while guaranteeing $\mathcal{L}_2$ convergence and an $H_\infty$
performance of the distributed attack detector, remains unchanged we
propose a different technique which suits time-varying problems such as the one
considered here. The idea of this technique is to decentralize the design
process (but not the detectors themselves) to make it tractable.  

To explain the decentralized design of the attack detector, let
$z_i=e_i-\hat{e}_i$, $\delta_i=\epsilon_i-\hat\epsilon_i$ be estimation
errors of the observer  (\ref{ext.obs.nu.1.om}). It is easy to see from (\ref{ext.obs.nu.1.om}) that
these errors evolve according to the equations
\begin{eqnarray}
    \dot{z}_i&=&(A(t) - \hat L_i(t)C_i(t)-\sum_{j\in
      \mathbf{N}_i}\hat K_{ij}(t)W_{ij})z_i - F_i\Upsilon_i\delta_i
    \nonumber \\ && - \sum_{j\in
      \mathbf{N}_i}\hat K_{ij}(t)(W_{ij}z_j+H_{ij}v_{ij}) + Bw \nonumber \\ 
&& -\hat L_i(t)D_i(t)v_i 
      +F_i\nu_i, \nonumber \\
\dot \delta_i&=&\Omega_i \delta_i-\check L_i(t)C_i(t)z_i-\sum_{j\in
      \mathbf{N}_i}\check K_{ij}(t)W_{ij}z_i +\Gamma_i \nu_i\nonumber \\
&& - \sum_{j\in
      \mathbf{N}_i}\check K_{ij}(t)(W_{ij}z_j+ H_{ij}v_{ij}) 
\label{ext.error.0} \label{ext.error} \\
&&  z_i(0)=x_0-\xi_i, \quad \delta_i(0)=0. \nonumber 
\end{eqnarray}
Here we used the notation $\hat L_i=L_i+\bar L_i$, $\hat
K_{ij}=K_{ij}+\bar K_i$. Also, let us introduce
signals
\begin{equation}
  \label{eta}
\eta_{ij}=W_{ij}z_j, \quad j\in\mathbf{N}_i. 
\end{equation}
Altogether, for each node $i$, $q_i$ such signals are introduced, 
where
$q_i$ is the cardinality of the set $\mathbf{N}_i$, also known as in-degree
of node $i$. Each signal $\eta_{ij}$ is of dimension $p_{ij}$, i.e., it
matches the dimension of the corresponding communication signal $c_{ij}$.
The signals $\eta_{ij}$ serve as interconnection signals for the
large-scale system (\ref{ext.error}). Then the error system
(\ref{ext.error}) can be written as 
\begin{eqnarray}
  \label{err.combined}
\frac{d}{dt}\lambda_i&=&(\mathbf{A}_i(t)-\mathbf{L}_i(t)\mathbf{C}_i(t))\lambda_i+ \mathbf{B}(t)\left[\begin{array}{c}w \\ \nu_i
  \end{array}\right] \nonumber \\
&& 
-\mathbf{L}_i(t)\mathbf{D}_i(t)\left[\begin{array}{c}v_i \\ \bar v_i \\
    \bar \eta_i \end{array}\right],
\end{eqnarray}
where we used the following notation:
\allowdisplaybreaks
\begin{eqnarray}
\lambda_i&=&\left[\begin{array}{c}z_i\\\delta_i
  \end{array}\right], \quad
\bar v_i= \left[\begin{array}{ccc} v_{ij_1}' & \ldots &
    v_{ij_{q_i}}'\end{array}\right]', \nonumber \\
\bar \eta_i&=& \left[\begin{array}{ccc} (Z^{-1/2}_{ij_1}\eta_{ij_1})' &
    \ldots &  (Z^{-1/2}_{ij_{q_i}}\eta_{ij_{q_i}})'  
  \end{array}\right]', \nonumber \\
\mathbf{A}_i(t)&=&\left[\begin{array}{cc}
A(t) & -F_i\Upsilon_i\\
0 & \Omega_i
    \end{array}
  \right], \quad 
\mathbf{B}_i=\left[\begin{array}{cc}
B(t) & F_i\\
0 & \Gamma_i
    \end{array}
  \right], \nonumber \\
\mathbf{C}_i(t)&=&\left[\begin{array}{cc}
C_i(t) & 0 \\
W_{ij_1} & 0 \\
\vdots & \vdots \\
W_{ij_{q_i}} & 0
    \end{array}
  \right], \quad 
\mathbf{L}_i=\left[\begin{array}{cccccc}
\hat L_i & \hat K_{ij_1} & \ldots & \hat K_{ij_{q_i}} \\ 
\check L_i & \check K_{ij_1} & \ldots & \check K_{ij_{q_i}}
\end{array}\right],
\label{Ldef}
\\
\mathbf{D_i}(t) &=& \left[\begin{array}{ccccccc} D_i(t) & 0 & \ldots & 0 & 0 & \ldots & 0  \\
                                 0   & H_{ij_1} & \ldots & 0 & Z_{ij_1}^{1/2}&
                                 \ldots & 0 \\ 
                                  \vdots & \vdots & \ddots & \vdots &
                                  \vdots & \ddots & \vdots \\
                                  0   & 0 & \ldots  & H_{ij_{q_i}} & 0 &
                                 \ldots & Z_{ij_{q_i}}^{1/2} 
                                      \end{array}\right]. 
\nonumber
\end{eqnarray}
Here $Z_{ij}$ $i=1,\ldots, N$, $j\in \mathbf{N}_i$ are certain square
$p_{ij}\times p_{ij}$ positive definite matrices. It will be assumed that
each matrix  $\mathbf{E}_i(t)=\mathbf{D_i}(t)\mathbf{D_i}'(t)$ is
positive definite for all $t$.

The system (\ref{err.combined}) can be regarded as an uncertain system 
affected by the disturbances $w$, $v_i$, $v_{ij}$ and signals $\nu_i$,
$Z_{ij}^{-1/2}\eta_{ij}$. The latter signals will be treated as
fictitious disturbances. This allows us to apply the $H_\infty$
filtering theory to obtain a gain coefficient $\mathbf{L_i}(t)$ which
guarantees that the impact of the disturbances 
on the errors $z_i$, $\delta_i$  is attenuated. 


Introduce collection of positive definite $(n+n_{f_i})\times (n+n_{f_i})$
block-diagonal matrices 
$\mathbf{R}_i,\mathbf{X}_i$, $i=1\ldots,N$, partitioned as
\[
\mathbf{R}_i=\left[\begin{array}{cc}R_i & 0\\ 0 & \check R_i
  \end{array}\right],
\quad 
\mathbf{X}_i=\left[\begin{array}{cc}X_i & 0\\ 0 & \check X_i
  \end{array}\right],
  \] 
with $n\times n$ matrices $R_i$, $X_i$ and $n_{f_i}\times n_{f_i}$ matrices
$\check R_i$, $\check X_i$. In addition, define the matrix
\begin{eqnarray}
  \Phi=[\Phi_{ij}]_{i,j=1}^N, 
\end{eqnarray}
composed of the blocks
\[
\Phi_{ij}=\begin{cases}\Delta_i, & i=j,\\
-W_{ij}'U_{ij}^{-1}W_{ij}, & i\neq j,~j\in\mathbf{N}_i,\\
0 & i\neq j,~j\not\in\mathbf{N}_i,
\end{cases}
\]
where 
\begin{eqnarray*}
U_{ij}=H_{ij}H_{ij}'+Z_{ij},\quad
\Delta_i=\sum_{j\in\mathbf{N}_i}W_{ij}'U_{ij}^{-1}Z_{ij}U_{ij}^{-1}W_{ij}.
\end{eqnarray*}
Also, let
\[
R=\mathrm{diag}[R_1, \ldots, R_N], \quad 
\Delta=\mathrm{diag}[\Delta_1, \ldots, \Delta_N]. 
\]
\begin{theorem}\label{T1}
Suppose there exists a constant $\gamma>0$ and positive definite
symmetric matrices $R_i$, $\check R_i$, $Z_{ij}$, $j\in \mathbf{N}_i$,
$i=1, \ldots N$, such that  
\begin{enumerate}[(i)]
\item
the following linear matrix inequalities are satisfied
\begin{eqnarray}
\label{LMI}
&&  R+\gamma^2(\Phi+\Phi'-\Delta) > I, \\
&&  \check R_i>I; \label{LMI.check}
\end{eqnarray}
\item
each differential Riccati equation  
\begin{eqnarray}\label{Riccati} 
\dot{\mathbf{Y}}_i &=& \mathbf{A}_i\mathbf{Y}_i+\mathbf{Y}_i\mathbf{A}_i'
\nonumber \\
&& - \mathbf{Y}_i(\mathbf{C}_i'\mathbf{E}_i^{-1}\mathbf{C}_i
-\frac{1}{\gamma^2}\mathbf{R}_i)\mathbf{Y}_i
+\mathbf{B}_i\mathbf{B}_i', \qquad  \\
\mathbf{Y}_i(0)&=&\mathbf{X}_i^{-1}, \nonumber  
\end{eqnarray}
has a positive definite symmetric bounded solution $\mathbf{Y}_i(t)$ on the
interval $[0,\infty)$, i.e., for all $t\ge 0$,
$\alpha_1I<\mathbf{Y}_i(t)=\mathbf{Y}_i'(t)<\alpha_2I$, for some $\alpha_{1,2}>0$.  
\end{enumerate}
Then the network of observers (\ref{ext.obs.nu.1.om}) with 
the coefficients $\bar L_i$, $\bar K_{ij}$, $\check L_i$, $\check K_{ij}$,
obtained by partitioning the matrices
\begin{eqnarray}
  \label{L}
  \mathbf{L}_i(t)=\mathbf{Y}_i(t)\mathbf{C}_i(t)'\mathbf{E}_i^{-1}(t).
\end{eqnarray}
according to (\ref{Ldef}), guarantees that the noise- and attack-free system
(\ref{ext.error}) is exponentially stable, while in the presence of
disturbances or an attack it holds that 
\begin{eqnarray*}
\int_0^{+\infty}\|\varphi_i-f_i\|^2dt <\infty \quad \forall i.
\end{eqnarray*}
\end{theorem}

The proof of the theorem is removed for brevity.

\begin{remark}
It follows from Theorem~\ref{T1} that $z_i\in \mathcal{L}_2[0,\infty)$,
therefore every signal $\eta_{ij}$ defined in (\ref{eta}) is
$\mathcal{L}_2$-integrable. This allows us to conclude that every
observer~(\ref{err.combined}) has a local disturbance attenuation
property. Specifically, according to the $H_\infty$ filtering
theory~\cite{Basar-Bernhard} it follows from the condition (ii) of
Theorem~\ref{T1} that
\begin{eqnarray}
\label{Hinf.property.ii}
\lefteqn{\int_0^T(z_i'R_iz_i+\delta_i'\check R_i \delta_i)dt} && \nonumber \\ 
&\le& \gamma^2\bigg(
\Vert x_0-\xi_i \Vert_{X_i}^ 2 + \int_0^T\Big[ \Vert w \Vert^2  
+\Vert v_i\Vert^2 \nonumber \\
&&+ \sum_{j\in\mathbf{N}_i}(\Vert v_{ij}\Vert^2
+ \Vert W_{ij}z_j \Vert^2_{Z^{-1}_{ij}})\Big]dt\bigg)  
\end{eqnarray}
This condition reveals the role of the matrices $R_i$, $\check
R_i$ and $Z_{ij}$ included in the Riccati equation (\ref{Riccati}) as
design parameters. These matrices weigh the output error of the observer
(\ref{ext.error}) at node $i$ against the information 
about the errors at the neighbouring nodes which supply information to node
$i$. 
\end{remark}

\section{Conclusion} \label{sec:conclusion}
The paper has proposed a decentralized $H_\infty$ synthesis method for the
design of distributed observers for detecting biasing attacks on
distributed filter networks. The proposed detectors can pick a biasing
attack from local sensory information complemented with information
extracted form the routine information exchange within the network. Our
method accounts for the fact that such filters operate in noisy
environments, therefore $H_\infty$ performance of the 
proposed detectors against disturbances is also guaranteed.

\newcommand{\noopsort}[1]{} \newcommand{\printfirst}[2]{#1}
  \newcommand{\singleletter}[1]{#1} \newcommand{\switchargs}[2]{#2#1}

\end{document}